\documentstyle[epsfig,mathptm]{mn}
\DeclareMathVersion{bold}
\newif\ifAMStwofonts
\AMStwofontstrue                
\DeclareMathVersion{bold}                    
\DeclareMathAlphabet{\mathbfit}{OT1}{cmr}{bx}{it}
\SetMathAlphabet\mathbfit{bold}{OT1}{cmr}{bx}{it}
\DeclareMathAlphabet{\mathbfss}{OT1}{cmss}{bx}{n}
\SetMathAlphabet\mathbfss{bold}{OT1}{cmss}{bx}{n}
\ifAMStwofonts
  \ifCUPmtlplainloaded \else
    \DeclareSymbolFont{UPM}{U}{eur}{m}{n}
    \SetSymbolFont{UPM}{bold}{U}{eur}{b}{n}
    \DeclareSymbolFont{AMSa}{U}{msa}{m}{n}
    \DeclareMathSymbol{\upi}{0}{UPM}{"19}
    \DeclareMathSymbol{\umu}{0}{UPM}{"16}
    \DeclareMathSymbol{\upartial}{0}{UPM}{"40}
    \DeclareMathSymbol{\leqslant}{3}{AMSa}{"36}
    \DeclareMathSymbol{\geqslant}{3}{AMSa}{"3E}

    \let\leq=\leqslant 
    \let\geq=\geqslant 
  \fi
\fi
\title[Wavelets and non-Gaussianity in the COBE data]
{Do wavelets really detect non-Gaussianity in the 4-year COBE data?}
\author[P. Mukherjee et al.]
{P. Mukherjee, M.P. Hobson and A.N.~Lasenby\\
Astrophysics Group, Cavendish Laboratory, Madingley Road, 
Cambridge CB3 OHE, UK\\}
\date{Accepted ???. Received ???; in original form 20 September 1999}
\pagerange{\pageref{firstpage}--\pageref{lastpage}}
\pubyear{1998} 

\begin{document} 
\maketitle 
\label{firstpage}

\begin{abstract}
We investigate the detection of non-Gaussianity in the 4-year COBE
data reported by Pando, Valls-Gabaud \& Fang (1998), using a technique
based on the discrete wavelet transform. Their analysis was
performed on the two DMR faces centred on the North and South Galactic
poles respectively, using the Daubechies 4 wavelet basis. We show that
these results depend critically on the orientation of the data, and
so should be treated with caution. For two distinct orientations of
the data, we calculate unbiased estimates of the skewness, kurtosis
and scale-scale correlation of the corresponding
wavelet coefficients in all of the available scale domains of the
transform.  We obtain several detections of non-Gaussianity 
in the DMR-DSMB map at greater
than the 99 per cent confidence level, but most of these occur on
pixel-pixel scales and are therefore not cosmological in
origin. Indeed, after removing all multipoles beyond $\ell = 40$
from the COBE maps, only one robust detection remains. Moreover, using
Monte-Carlo simulations, we find that the probability of obtaining
such a detection by chance is 0.59. We repeat the analysis for
the 53+90 GHz coadded COBE map. In this case, after removing
$\ell > 40$ multipoles, two non-Gaussian detections 
at the 99 per cent level remain.
Nevertheless, again using Monte-Carlo simulations, we find that the
 probability of obtaining two such detections by chance is 0.28. 
Thus, we conclude the wavelet technique does
{\em not} yield strong evidence for non-Gaussianity of cosmological
origin in the 4-year COBE data.
\end{abstract}


\section{Introduction}
\label{intro}

Observations of temperature fluctuations in the cosmic microwave
background (CMB) provide a valuable means of distinguishing between two
competing theories for the formation of structure in the
early Universe. Currently, the most favoured theory is the simple
inflationary cold-dark-matter (CDM) model, for which the
distribution of temperature fluctuations in the CMB should be
Gaussian. The second class of theories invokes the formation of
topological defects such as cosmic strings, monopoles or textures,
which should imprint some non-Gaussian features in the CMB (Bouchet,
Bennett \& Stebbins 1988; Turok 1996). Thus, the detection (or
otherwise) of a non-Gaussian signal in the CMB is an important means
of discriminating between these two classes of theory.

In order to test for large-scale non-Gaussianity in the CMB, the
4-year COBE-DMR dataset (in various forms) has already been analysed
using a number of different statistical techniques, as discussed
below.  These tests have been performed either on some combination of
the 31-, 53- and 90-GHz A \& B 4-year DMR maps, or the 4-year DMR maps
from which Galactic emission has been removed.  Two such
Galaxy-removed maps are generally available, each one created using a
different separation method.  The DMR-DCMB map is a linear combination
of all six individual COBE-DMR maps designed to cancel the free-free
emission (Bennett et al. 1992), whereas the DMR-DSMB map is
constructed by first subtracting templates of synchrotron and dust
emission and then removing free-free emission (Bennett et
al. 1994).

The first investigation of non-Gaussianity in the 4-year COBE data was
performed by Kogut et al. (1996). This analysis used the 4-year DMR 53
GHz $(A+B)/2$ map at high latitudes ($|b| > 20\degr$) with cut-outs
near Ophiuchus and Orion (Bennett et al. 1996), and found that
traditional statistics such as the three-point correlation function,
the genus and the extrema correlation function, were completely
consistent with a Gaussian CMB signal.  Colley, Gott \& Park (1996)
also computed the genus statistic, but for the DMR-DCMB map with $|b|
> 30\degr$, and arrived at similar conclusions. The full set of
Minkowski functionals were computed for the 4-year 53 GHz $(A+B)/2$
map (with a smoothed Galactic cut) by Schmalzing \& Gorski (1998),
taking proper account of the curvature of the celestial sphere. They
also concluded that the CMB is consistent with a Gaussian random field
on degree scales. On computing the bi-spectrum of the 4-year COBE
data, Heavens (1998) also found no evidence for non-Gaussianity.
Finally, Novikov, Feldman \& Shandarin (1999) have calculated the
partial Minkowski functionals for both the DMR-DCMB and DMR-DSMB maps
and do report detections of non-Gaussianity, but the analysis
was performed without making a Galactic cut and the detections most
probably result from residual Galactic contamination.

Recently, however, two apparently robust detections of
non-Gaussianity in the 4-year COBE data have been reported.
Ferreira, Magueijo \& Gorski (1998) applied a technique based on the normalised
bi-spectrum to a map created by averaging the 53A, 53B, 90A and 90B
4-year COBE-DMR channels (each weighted according to the inverse of
its noise variance) and then applying the extended Galactic cut of
Banday et al. (1997) and Bennett et al. (1996). They concluded that
Gaussianity can be rejected at the 98 per cent confidence level, with
the dominant non-Gaussian signal concentrated near the multipole $\ell
=16$. This non-Gaussian signal is certainly present in the 
COBE data, but Banday, Zaroubi \& Gorski (1999) have now shown that it is not
cosmological in origin and is most likely the result of an
observational artefact. Nevertheless, using an extended bi-spectrum
analysis, Magueijo (1999) reports a new non-Gaussian
signal above the 97 per cent level, even after removing
the observational artefacts discovered by Banday et al.

A second detection of non-Gaussianity was reported by Pando,
Valls-Gabaud \& Fang (1998) (hereinafter PVF), who applied a
technique based on the discrete wavelet transform (DWT) to Face 0 and
Face 5 of the QuadCube pixelisation of the DMR-DCMB and DMR-DSMB maps
in Galactic coordinates (i.e. the North and South Galactic pole
regions respectively).  PVF computed the skewness, kurtosis and
scale-scale correlation of the wavelet coefficients of DMR maps in
certain domains of the wavelet transform, and compared these
statistics with the corresponding probability distributions computed from
1000 realisations of simulated COBE observations of a Gaussian CMB
sky. In all cases, they found that the skewness and kurtosis of the
wavelet coefficients were consistent with a Gaussian CMB signal. On
the other hand, the scale-scale correlation coefficients showed
evidence for non-Gaussianity at the 99 per cent confidence level on
scales of 11--22 degrees in Face 0 of both the DMR-DCMB and DMR-DSMB
maps. Nevertheless, in both maps, Face 5 was found to be consistent
with Gaussianity. We note that Bromley \& Tegmark (1999) confirm the
findings of both PVF and Ferreira et al. (1998).

In this paper, we also apply to the 4-year COBE data a
non-Gaussianity test based on the skewness, kurtosis and scale-scale
correlation of the wavelet coefficients. In the analysis presented
below, however, we calculate the skewness and kurtosis statistics
using {\em unbiased} estimators based on $k$-statistics (Hobson,
Jones \& Lasenby 1999 - hereinafter HJL), as opposed to the
straightforward calculation of sample moments employed by PVF. For
the scale-scale correlation, we adopt the same definition as that used
by PVF. We also note that the analysis presented below is slightly 
more general than
that presented by PVF, since we calculate the statistics of the
wavelet coefficients in {\em all} the available domains of the wavelet
transform, as opposed to using only those regions that represent
structure in the maps on the same scale in the horizontal and vertical
directions. 

Perhaps the most important point addressed in the analysis
presented here, however, is the fact that non-Gaussianity tests based
on any orthogonal compactly-supported 
wavelet decomposition are sensitive to the orientation
of the input map. This is discussed in detail below. As an
illustration of this point, we therefore present the results of two 
separate analyses, in which the relative orientations of the input
maps differ by 180 degrees.
Nevertheless, it should be remembered that, in general, different
techniques for detecting non-Gaussianity are each sensitive to different
ways in which the data may be non-Gaussian. 
We should therefore not be too surprised if the detailed results of an
analysis are orientation dependent. Obviously, it would be troubling
if the general conclusions concerning non-Gaussianity of the data
depended on orientation, but that is not the case here.

\section{The wavelet decomposition}

The basics of the wavelet non-Gaussianity test are discussed in detail
in HJL and also by PVF and so we give only a brief outline
here. The two-dimensional discrete wavelet transform (DWT) (Daubechies
1992, Press et al. 1994) performs the decomposition of a planar
digitised image of size $2^{J_1}\times 2^{J_2}$ into the sum of a set
of two-dimensional planar (digitised) wavelet basis functions
\begin{equation}
\frac{\Delta T}{T}(\bmath{x}_i)
= \sum_{j_1=0}^{J_1-1}\sum_{j_2=0}^{J_2-1}
  \sum_{l_1=0}^{2^{j_1}-1}\sum_{l_2=0}^{2^{j_2}-1}
  b_{j_1,j_2;l_1,l_2} \psi_{j_1,j_2;l_1,l_2}(\bmath{x}_i).
\label{dwtdef}
\end{equation}
In equation (\ref{dwtdef}), the wavelets
$\psi_{j_1,j_2;l_1,l_2}(\bmath{x})$ (with $j_1,j_2,l_1,l_2$ taking the
values indicated in the summations) form a complete and orthogonal set
of basis functions. Each two-dimensional wavelet is simply the direct
tensor product of the corresponding one-dimensional wavelets
$\psi_{j_1;l_1}(x)$ and $\psi_{j_2;l_2}(y)$, which in turn are defined
in terms of the dilations and translations of some mother wavelet
$\psi(x)$ via
\begin{equation}
\psi_{j_1;l_1}(x) = \left(\frac{2^{j_1}}{L}\right)^{1/2}\psi(2^{j_1} x/L-l_1),
\end{equation}
where $0 \leq x \leq L$, and a similar expression holds for
$\psi_{j_2;l_2}(y)$. Thus, the scale indices $j_1$ and $j_2$
correspond to the scales $L/2^{j_1}$ and $L/2^{j_2}$ in the $x$- and
$y$-directions respectively (so $J_1$ and $J_2$ are the smallest
possible scales -- i.e. one pixel -- in each direction), whereas the
location indices $l_1$ and $l_2$ correspond to the $(x,y)$-position
$(Ll_1/2^{j_1},Ll_2/2^{j_2})$ in the image.  Since each wavelet basis
function $\psi_{j_1,j_2;l_1,l_2}(x,y)$ is localised at the
relevant scale/position, the corresponding wavelet coefficient
$b_{j_1,j_2;l_1,l_2}$ measures the amount of signal in the image at
this scale and position.

\subsection{Orientation sensitivity}
\label{orient}

At this point, it is important to note the sensitivity of
the orthogonal wavelet decomposition to the orientation of the original input
map. As shown by Daubechies (1992), it is impossible to construct an
orthogonal wavelet basis, in which the basis functions are both
symmetric (or anti-symmetric) 
and have compact support. This asymmetry of the basis
functions is the cause of the orientation sensitivity. This is most
easily appreciated by considering an input map consisting of just one
of the wavelet basis functions. If this map is rotated through 
180 degrees (say), then because the basis functions are asymmetric it
is not possible to represent the rotated basis functions in terms of just one
of the original basis function. Instead, the signal in the rotated map
must be represented by several wavelet basis functions with different
scale and position indices. Thus any statistics based on the wavelet
coefficients are sensitive to the orientation of the original input
map. Since the origin of this effect is the asymmetry of the
one-dimensional wavelet basis functions, it also occurs for two-dimensional
orthogonal wavelet decompositions based on the Mallat algorithm
(Mallat 1989), which is also commonly called the multiresolution
analysis method. In order to obtain wavelet statistics that are invariant
under 90, 180, 270 degrees rotations of the input image (and also
insensitive to cyclic translations of the image by an arbitrary number
of pixels in each direction), it is necessary to use the {\em \`a
trous} wavelet algorithm (see e.g. Starck, Murtagh \& Bijaoui 1998) 
with a symmetric filter function. The
application of this technique to the detection of non-Gaussianity in
the CMB will be presented in a forthcoming paper.

\begin{figure}
\centerline{\epsfig{
file=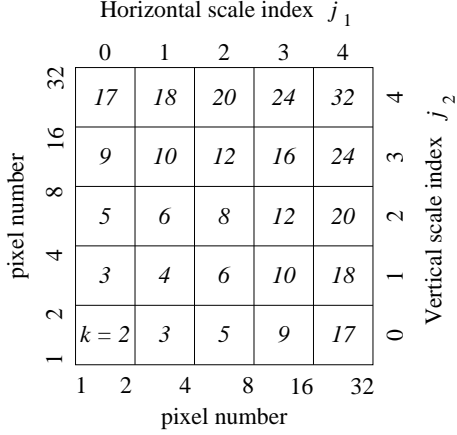,width=6cm}}
\caption{The regions of the two-dimensional wavelet domain for a
$32\times 32$ image. The italic numbers show the value of
$k=2^{j_1}+2^{j_2}$ in each region; see text for details.}
\label{domains}
\end{figure}

\subsection{Application to COBE data}

In this paper, we will be concerned with Face 0 and Face 5 of the
COBE QuadCube pixelisation scheme in Galactic coordinates, 
each of which consists of $32
\times 32$ equal-area pixels (i.e. $J_1=J_2=5$) of size $(2.8
\degr)^2$. Thus the scale $j$ corresponds to an angular scale of $2.8
\times 2^{4-j}$.  Following the discussion by HJL, the structure of
the corresponding wavelet domain is shown in Fig.~\ref{domains}, where
the pixel numbers are plotted on a logarithmic scale. We see that the
domain is partitioned into separate regions according to the scale
indices $j_1$ and $j_2$ in the horizontal and vertical directions
respectively.  
Thus regions with $j_1=j_2$ contain wavelets basis
functions that represent the image at the same scale in the two
directions, whereas regions with $j_1 \neq j_2$ describe the image on
different scales in the two directions.  
As discussed in HJL, regions with $j_1=0$ or $j_2=0$ actually
contain basis functions that are tensor products of different
one-dimensional basis and so for the remainder of this paper we
restrict our attention to regions with $j_1,j_2 \geq 1$.
We also define the integer
variable $k=2^{j_1}+2^{j_2}$, which serves as a measure of inverse
scale length, and is constant within each region of the wavelet
domain.  We note that the value of $k$ is not altered if the values
$j_1$ and $j_2$ are interchanged. In this paper, we also restrict
ourselves to the Daubechies 4 wavelet basis used by PVF, although 
analogous analyses may also be 
performed for other orthogonal discrete wavelet bases, and indeed
similar results to those presented in Section 3 are obtained.

\subsection{Skewness and kurtosis spectra}

Following HJL, when considering the statistics of the wavelet
coefficients $b_{j_1,j_2;l_1,l_2}$ of an image, it is useful to
consider separately all those coefficients that share each value of
$k$. For each value of $k$, we then use the corresponding wavelet
coefficients to calculate estimators of the skewness $\hat{S}$ and
(excess) kurtosis $\hat{K}$ of the parent distribution from which
the coefficients were drawn. We therefore obtain the skewness and
kurtosis `spectra' $\hat{S}(k)$ and $\hat{K}(k)$ for the image.  

As mentioned in the Introduction, at this point our method diverges
from that used by PVF in two ways. Firstly, PVF only consider
regions of the wavelet domain for which $j_1=j_2$ and $j_1,j_2 \geq 1$
(i.e. $k=4,8,16,32$), whereas we consider all regions with $j_1,j_2
\geq 1$.  Secondly, we calculate the estimators $\hat{S}$ and
$\hat{K}$ in a different way from that adopted in PVF, as follows.
At each value of $k$ the skewness and (excess) kurtosis of the parent
distribution of the wavelet coefficients are given by
\begin{eqnarray}
S & = & \mu_3/\mu_2^{3/2} = \kappa_3/\kappa_2^{3/2},
\label{skewdef} \\
K & = & \mu_4/\mu_2^2 - 3 = \kappa_4/\kappa_2^2,
\label{kurtdef}
\end{eqnarray}
where $\mu_n$ is the $n$th central moment of the distribution and
$\kappa_n$ is the $n$th cumulant (see HJL for a brief discussion).
In PVF, the estimators $\hat{\mu}_n$ of the central moments are
simply taken to be the central moments of the sample of wavelet
coefficients. It is easily shown, however, that these estimators are
biased, so that $\langle \hat{\mu}_n \rangle \neq \mu_n$, and this
bias is quite pronounced when the sample size is small (as it is in
this case). PVF then estimate the skewness and (excess) kurtosis by
inserting the biased estimators $\hat{\mu}_n$ into (\ref{skewdef})
and (\ref{kurtdef}) respectively.  Thus, the corresponding estimators
$\hat{S}$ and $\hat{K}$ are also significantly biased.
In this paper, we instead calculate our estimates of the
skewness and (excess) kurtosis using $k$-statistics (see Kenney \&
Keeping 1954; Stuart \& Ord 1994; HJL).  These provide unbiased
estimates $\hat{\kappa}_n$ of the cumulants of the parent population
from which the wavelet coefficients were drawn. These unbiased
estimators of the cumulants are then inserted into (\ref{skewdef}) and
(\ref{kurtdef}) to obtain the estimators $\hat{S}$ and $\hat{K}$.

\subsection{Scale-scale correlation spectrum}

In addition to the skewness and kurtosis spectra, we may also
measure the correlation between the different domains
of the wavelet transform by defining the estimators of
the scale-scale correlation as 
\begin{equation}
\hat{C}^p_{j_1,j_2} = 
\frac{2^{j_1+j_2+2}
\sum_{l_1}\sum_{l_2}
b^p_{j_1,j_2;[l_1/2],[l_2/2]}b^p_{j_1+1,j_2+1;l_1,l_2}}
{\sum_{l_1}\sum_{l_2} b^p_{j_1,j_2;[l_1/2],[l_2/2]}
\sum_{l_1}\sum_{l_2} b^p_{j_1+1,j_2+1;l_1,l_2}}.
\label{ssdef}
\end{equation}
In equation (\ref{ssdef}), the sums on $l_1$ extend from
$0$ to $2^{j_1+1}-1$ (similarly for $l_2$), 
$p$ is an even integer and $[~]$ denotes the integer part.
Thus $C^p_{j_1,j_2}$ measures the correlation between the wavelet
coefficients in the domains $(j_1,j_2)$ and $(j_1+1,j_2+1)$.
In PVF, it was assumed that $j_1=j_2$, so that the correlation
of wavelet coefficients were only calculated between adjacent
diagonal domains in Fig.~\ref{domains}. When $j_1 \neq j_2$, however,
it is convenient to extend the sums in (\ref{ssdef}) to include
also the corresponding domains with $j_1$ and $j_2$ interchanged.
Thus, in each case, we in fact measure the correlation between 
wavelet coefficients with inverse scalelengths of $k$ and $2k$ 
respectively (see Fig.~\ref{domains}).
For each possible value of $k$, we denote this correlation
by $\hat{\cal C}^p(k)$, thereby producing a scale-scale correlation
spectrum. Following PVF,  we restrict our analysis
to the case where $p=2$.

\subsection{The non-Gaussianity test}

The skewness, (excess) kurtosis and scale-scale correlation
spectra $\hat{S}(k)$ , $\hat{K}(k)$  and $\hat{\cal C}^2(k)$
of the wavelet coefficients form the basis of the
non-Gaussianity test. The procedure is as follows. We first calculate
the $\hat{S}(k)$, $\hat{K}(k)$ and $\hat{\cal C}^2(k)$ 
spectra for Face 0 or Face 5 of the
4-year COBE map. We then generate 5000 realisations of an
all-sky CMB map drawn from an inflationary/CDM model 
with parameters $\Omega_{\rm m} = 1$,
$\Omega_{\Lambda} = 0$, $h=0.5$, $n=1$ and $Q_{\rm rms-ps}=18$ $\mu$K,
convolved with a $7\degr$-FWHM Gaussian beam.  For each realisation,
we then add random Gaussian pixel noise, where the rms of the noise in
each pixel is taken from the COBE rms noise map. The $\hat{S}(k)$,
$\hat{K}(k)$ and $\hat{\cal C}^2(k)$ 
spectra are then calculated for Face 0 and Face 5 of
each of the 5000 realisations to obtain approximate probability
distributions for the $\hat{S}(k)$, $\hat{K}(k)$ 
and $\hat{\cal C}^2(k)$ statistics when
the CMB signal is the chosen Gaussian inflationary/CDM model. By
comparing these probability distributions with the corresponding
spectra for Face 0 and Face 5 of 
the COBE map, we thus obtain (at each $k$-value) an estimate of the
probability that the CMB signal in the DMR-DSMB map is drawn from a
Gaussian ensemble characterised by the chosen inflationary/CDM model.
For each face, however, we obtain skewness and kurtosis
statistics at ten different $k$-values, and six different 
scale-scale correlation statistics. Thus, the total number of
statistics obtained for each face is 26, and care must be taken in
assessing the significance of non-Gaussianity detections at individual
$k$-values (see below). As discussed in section \ref{orient}, however,
the orthogonal wavelet decomposition is sensitive to the orientation
of the input map. Thus, we repeat the above non-Gaussianity test for
the case where Face 0 and Face 5 are both rotated through 180 degrees.

It is also clear that, to some extent, the results of such an analysis will
depend on the chosen parameters in the inflationary/CDM model via
the corresponding predicted ensemble-average power spectrum $C_\ell$,
from which the 5000 realisations are generated. Nevertheless, since
at each $k$-value the skewness and kurtosis statistics 
contain the variance $\mu_2$ of the wavelet coefficients
in their denominators, and the scale-scale correlation in (\ref{ssdef})
is similarly normalised, we would expect these statistics to be relatively
unaffected by changing the power spectrum of the inflationary/CDM
model. As an interesting test, we repeated our entire analysis 
for the case where the 5000 realisations were instead generated using
the maximum-likelihood $C_\ell$ spectrum calculated from the 4-year
COBE data by Gorski (1997). As expected, we found that the results
were virtually identical to those presented in the next Section.

\section{Results}

\subsection{The DMR-DSMB map}

In this Section, we present the results of the wavelet non-Gaussianity
test when applied to Face 0 and Face 5 of the 4-year COBE DMR-DSMB map
in Galactic coordinates. As mentioned in the Introduction, this
Galaxy-removed map is constructed by first subtracting templates of
synchrotron and dust emission and then removing the free-free emission
(Bennett et al. 1994). We find that the results of the non-Gaussianity
test are similar for both the DMR-DSMB and DMR-DCMB
Galaxy-removed maps.

\begin{figure*}
\centerline{\epsfig{
file=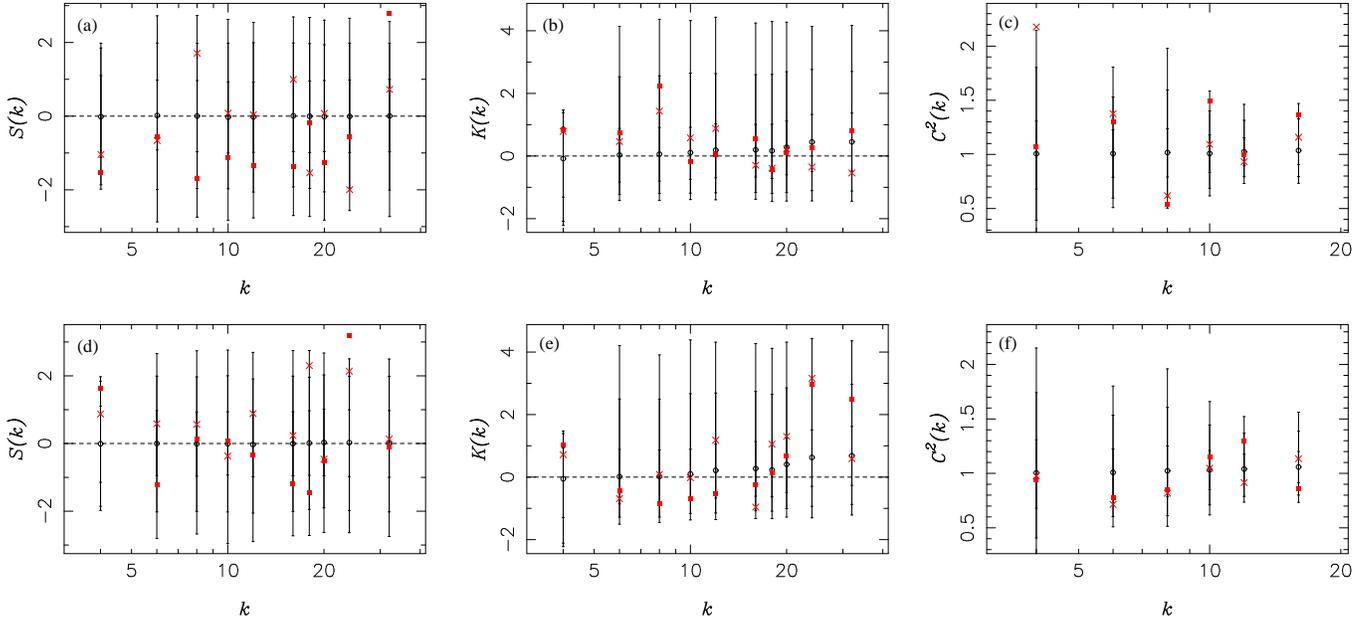,width=17.9cm}}
\caption{The $\hat{S}(k)$, $\hat{K}(k)$ and $\hat{\cal C}^2(k)$
spectra for Face 0 [plots (a), (b) and (c)] 
and Face 5 [plots (d), (e) and (f)]
of the DMR-DSMB map.
The crosses correspond to the orientation used by PVF (orientation A), whereas the
solid squares correspond to the map rotated through 180 degrees
(orientation B). 
The open
circles show the mean value for each statistic obtained from 5000
simulated COBE observations of CDM realisations, and the error bars
denote the 68, 95 and 99 per cent limits of the corresponding
distribution. For convenience, the $\hat{S}(k)$ and $\hat{K}(k)$
spectra have been normalised at each value of $k$ such that the
variance of the distribution obtained from the 5000 CDM realisations
is equal to unity.}
\label{dsmbf0}
\end{figure*}

The resulting $\hat{S}(k)$, $\hat{K}(k)$ 
and $\hat{\cal C}^2(k)$ spectra for
Face 0 and Face 5 of the DSMB map are plotted in 
Fig.~\ref{dsmbf0}. In each
plot, the crosses correspond to the values derived from the
DSMB map orientated in the same manner as that used by PVF
(orientation A), the solid 
squares correspond to the values obtained from the DSMB map after 
rotating it through 180 degrees (orientation B), 
and the open circles denote the mean 
of corresponding
distribution derived from the simulated COBE observations of the 5000
realisations of the inflationary/CDM model. The error bars denote the 
68, 95 and 99 per cent limits of the distributions. These
distributions were found to be virtually indistinguishable for the two
orientations of the COBE data. For convenience, 
the $\hat{S}(k)$ and $\hat{K}(k)$ spectra have
been normalised so that the variance of each distribution 
is equal to unity. Thus, for any particular $k$-value, a
estimate of the significance level can be read off directly from
the scale on the vertical axis. 

As mentioned above, we calculate the $\hat{S}(k)$ and $\hat{K}(k)$
spectra for all available domains of the wavelet transform, and the
$\hat{\cal C}^2(k)$ spectrum for all pairs of domains whose $k$-values
differ by a factor of 2 (with $j_1,j_2 \geq 1$ in each case; see
Fig.~\ref{domains}).  In contrast, PVF only considered domains with
$j_1=j_2$ and thus only obtained $\hat{S}(k)$ and $\hat{K}(k)$ values
for $k=4,8,16,32$, and $\hat{\cal C}^2(k)$ values at $k=4,8,16$.

We see from Fig.~\ref{dsmbf0} that for orientation A (crosses), all the 
points in the skewness and kurtosis spectra lie comfortably within their
respective Gaussian probability distributions for both faces. In the
scale-scale correlation spectrum, however, we confirm PVFs finding
of a point at $k=4$ that lies slightly outside the 99 per cent
confidence limit. On the other hand, for orientation B (solid squares)
we obtain two skewness detections someway 
beyond the 99 per cent 
confidence limit. These occur in Face 0 at $k=32$ and in Face 5 at $k=24$.
From Fig.~\ref{domains}, however, we see that these $k$-values
correspond to wavelet basis functions on small scales, corresponding
to pixel-to-pixel variations in the COBE map. Thus it is unlikely that this
non-Gaussianity is cosmological in origin; we return to this point below.
The kurtosis spectrum and scale-scale correlation spectra show no strong
non-Gaussian outliers for this orientation.

\begin{figure*}
\centerline{\epsfig{
file=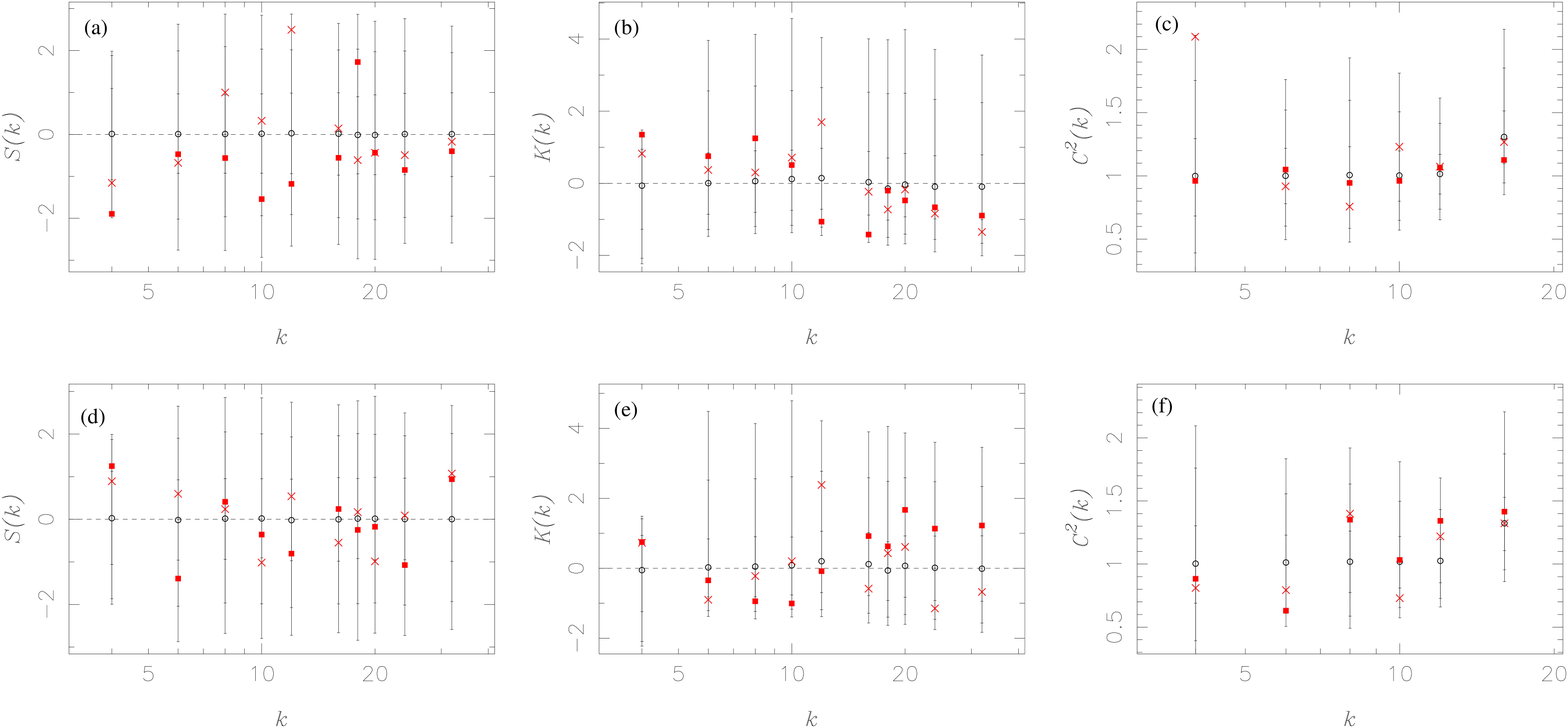,width=17.9cm}}
\caption{As in Fig.~\ref{dsmbf0}, but for the DMR-DSMB map
with all mutlipoles above $\ell=40$ removed.}
\label{dsmbf0filt}
\end{figure*}

In order to investigate the robustness of the high-$k$ outliers in the
$\hat{S}(k)$ spectra for orientation B, we repeated the analysis for 
the DSMB map
with all multipoles above $\ell=40$ removed. A similar filtering
process was also performed on each of the 5000 CDM
realisations. Since the 7-degree FWHM COBE beam essentially filters
out all modes beyond $\ell=40$, we would expect these modes to 
contain no contribution from the sky and consist
only of instrumental noise or observational artefacts.
We also repeated the filtering process for orientation A.
The corresponding $\hat{S}(k)$, $\hat{K}(k)$ 
and $\hat{\cal C}^2(k)$ spectra for two orientations of
Face 0 and Face 5 of the
filtered DSMB map are plotted in
Fig.~\ref{dsmbf0filt}.
We see immediately that the
high-$k$ skewness detections that were present in the unfiltered map
have now disappeared. This suggests that the non-Gaussianity present
in the original DSMB map is not cosmological in origin, and is most likely
an artefact resulting from the algorithm used
to subtract Galactic emission. From Fig.~\ref{dsmbf0filt}(c),
we also note that the three points
that lay outside the 95 per cent limit in the $\hat{\cal C}^2(k)$
spectrum for Face 0 of the original DSMB map in orientation B (see
Fig.~\ref{dsmbf0}(c)) have all now been brought well within the
Gaussian error bars. Thus we find no strong evidence for
non-Gaussianity in the filtered DSMB map in orientation B.
For orientation A, however, as we might expect, the level of
significance 
of the $\hat{\cal C}^2$
detection at $k=4$ was only slightly reduced by the filtering process.

\subsection{The 53+90 GHz coadded map}

Since the above analysis suggests some non-Gaussianity on
pixel scales in the DSMB map, possibly introduced by the Galaxy
subtraction algorithm, we repeat the analysis for the inverse noise
variance weighted average of the 53A, 53B, 90A and 90B COBE DMR
channels. 

\begin{figure*}
\centerline{\epsfig{
file=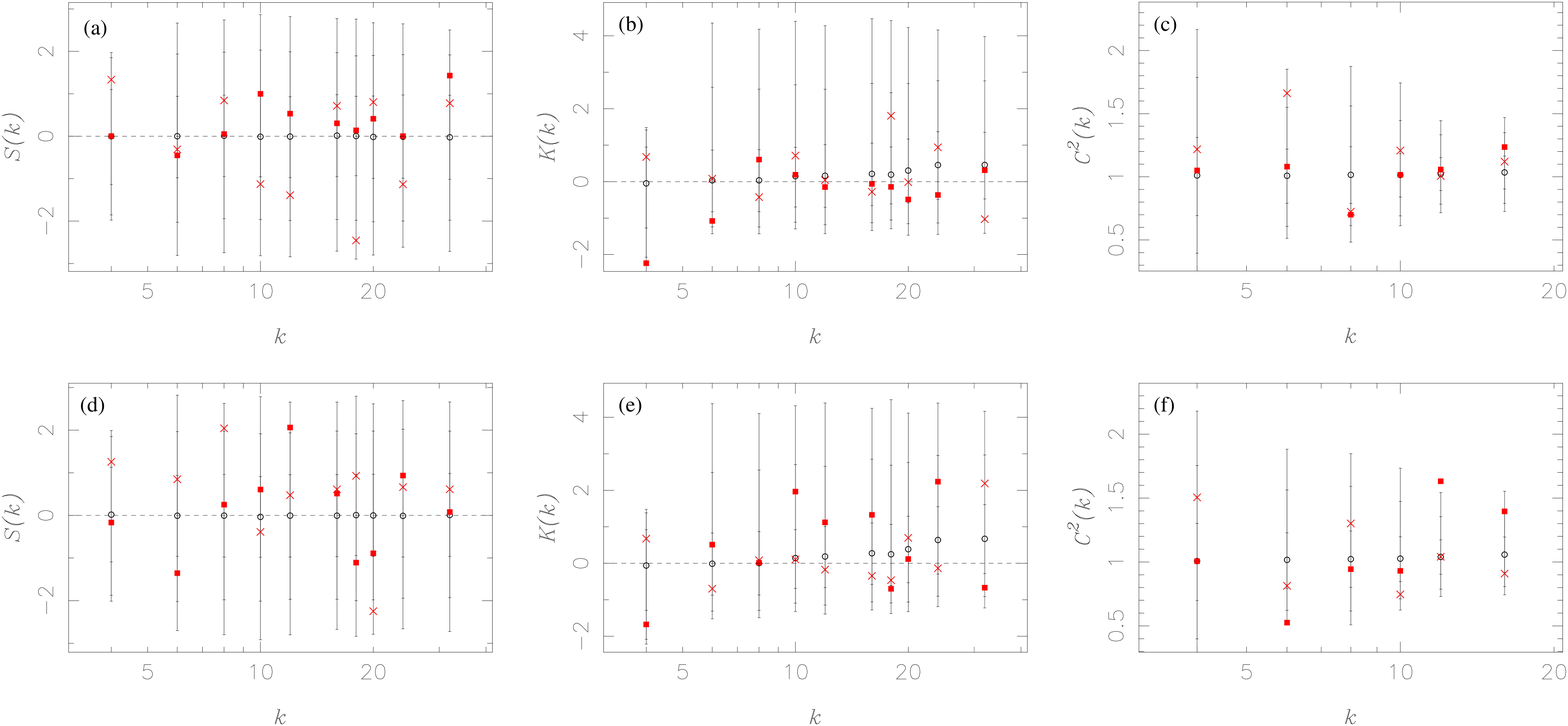,width=17.9cm}}
\caption{As in Fig.~\ref{dsmbf0}, but for the 53+90 GHz
coadded map.}
\label{coaddedf0}
\end{figure*}

Fig.~\ref{coaddedf0} shows the $\hat{S}(k)$, $\hat{K}(k)$ 
and $\hat{\cal C}^2(k)$ spectra for Face 0 and Face 5 of 
the $53+90$ GHz coadded map in both orientations.
For orientation A (crosses), none of the skewness, kurtosis or
scale-scale correlation statistics lies outside the corresponding 99
per cent limit. Also, for orientation B (solid squares), we see that,
in contrast to the DSMB map, no large detections of
non-Gaussianity are obtained at high $k$ in the skewness spectra. 
Nevertheless, outliers do occur at the 99 per cent level for Face 0
in the $\hat{K}(k)$ spectrum at $k=4$, and for Face 5 in the 
$\hat{\cal C}^2(k)$ at $k=6$ and $k=12$. Indeed, the last of these lies
someway outside the 99 per cent confidence limit.
However, this statistic measures the
correlation between the wavelet coefficients in the domains with
$k=12$ and $k=24$, and is therefore influenced primarily
by features in the map on the scale of one or two pixels in size.
\begin{figure*}
\centerline{\epsfig{
file=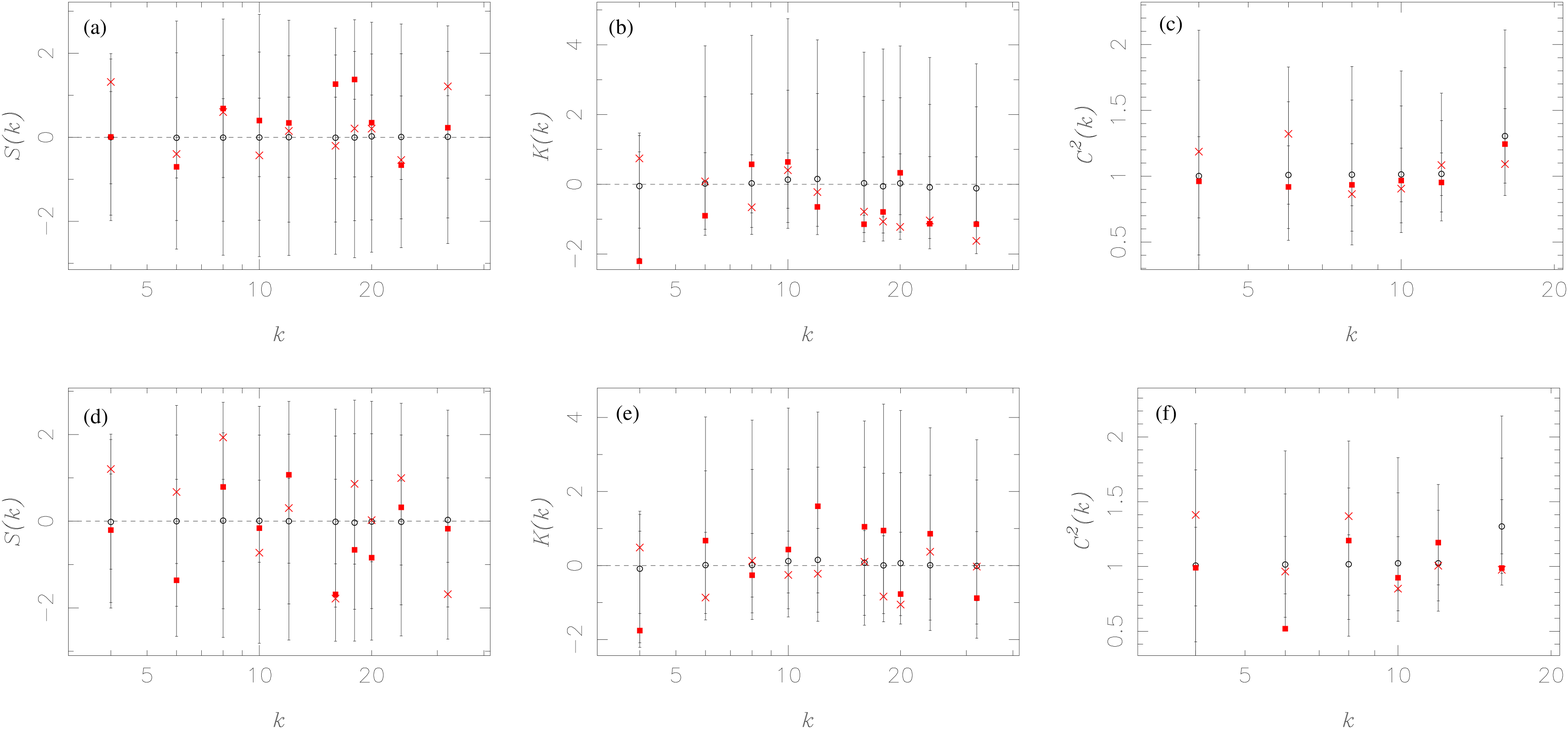,width=17.9cm}}
\caption{As in Fig.~\ref{dsmbf0}, but for the 53+90 GHz
coadded map with all multipoles above $\ell=40$ removed.}
\label{coaddedf0filt}
\end{figure*}

We once again tested the robustness of these putative detections of
non-Gaussianity by repeating the analysis after removing
all multipoles above $\ell=40$ from the COBE map and the CDM
realisations. The resulting spectra are shown in
Fig.~\ref{coaddedf0filt}. From
Fig.~\ref{coaddedf0filt}(f), we see that the large outlier
in $\hat{\cal C}^2(12)$ that was obtained for the
unfiltered map in orientation B has now reduced
to well within the Gaussian error bars. This suggests that the noise
in Face 5 of the coadded map may contain some non-Gaussian component.
Nevertheless, the two outliers
at the 99 per cent limit in $\hat{K}(4)$ for Face 0 and 
$\hat{\cal C}^2(6)$ for Face 5 in orientation B remain unaffected 
by the filtering process, and
thus might be interpreted as robust signatures of non-Gaussianity
on large scales. 

It is, however, important to remember that, although the significance
level is above the 99 per cent level for these individual statistics,
we must take into account the fact that no outliers are found in the
large number of other statistics we have calculated; this is discussed
below. It should also be
bourne in mind that no Galaxy
subtraction has been performed on the 53+90 GHz coadded map. Although,
our analysis is restricted to Face 0 and Face 5 of the COBE QuadCube,
which lie outside the standard Galactic cut, it is possible that
these faces may be contaminated to some extent by high-latitude Galactic
emission.

\section{Discussion and conclusions}

We have presented an orthogonal wavelet analysis of the 4-year COBE
data, in order to search for evidence of large-scale 
non-Gaussianity in the CMB. In particular, we 
identify an orientation sensitivity 
associated with this method, which must be borne in mind when
assessing its results. 

We find that several statistics in the
$\hat{S}(k)$, $\hat{K}(k)$ and $\hat{\cal C}^2(k)$ spectra for the
COBE DSMB and 53+90 GHz coadded maps (in orientations A and B)
lay outside the 99 per cent limit
of the corresponding probability distributions derived from 5000
simulated COBE observations of CDM realisations.  However, only one 
such outlier in the DSMB map and two outliers in the
53+90 GHz coadded map were found to be robust to the removal of 
all multipoles
above $\ell=40$ in the COBE map and CDM realisations. 
In the DSMB map, this occurs in $\hat{\cal C}^2(4)$ for Face 0 in
orientation A, and in the 53+90 GHz coadded COBE map the outliers
are in $\hat{K}(4)$ for Face 0 and $\hat{\cal C}^2(6)$ for Face 5, 
both in orientation B. 

We must, however, take care in assessing the significance of these
outliers. For each face and orientation
we calculate 26 different statistics. Thus for each data set
(either DSMB or 53+90 GHz coadded), the total number of statistics
is $2\times 2\times 26 = 104$,
and we must take proper account of the fact that a large number of
these show no evidence of non-Gaussianity (see, for example, Bromley
\& Tegmark 1999). Since the statistics presented here are not
independent of one another and generally do not possess Gaussian
one-point functions, the only way of obtaining a meaningful estimate
of the significance of our results is by Monte-Carlo
simulation. Indeed, in their bi-spectrum analysis of the 4-year COBE
data, Ferreira et al. (1998) used Monte-Carlo simulations and a
generalised $\chi^2$-statistic to assess their results.  In our case,
we adopt a slightly different approach and simply use the 5000 CDM
realisations to estimate the
probability of obtaining a given number of robust outliers at $> 99$
percent level in {\em any} of our 104 statistics, even when the
underlying CMB signal is Gaussian. For the DSMB data, we
obtained one robust outlier, and the corresponding probability of this
occuring by chance was found to be 0.59. For the 53+90 GHz coadded
data, two outliers were obtained, and the corresponding probability is
0.28. Therefore planar orthogonal wavelet analysis of the 4-year COBE 
data can only rule out
Gaussianity at the 41 per cent level in the DSMB data and at
the 72 per cent level in the 53+90 GHz coadded data. Thus, we conclude that
this method does {\em not} provide
strong evidence for non-Gaussianity in the CMB.

\section*{Acknowledgements}

The authors thank David Valls-Gabaud for his work in independently
verifying the numerical results presented here.  PM acknowledges
financial support from the Cambridge Commonwealth Trust. MPH thanks
the PPARC for financial support in the form of an Advanced Fellowship.

\bsp  
\label{lastpage}
\end{document}